\documentstyle[preprint,floats,aps]{revtex}

\newcommand{\doublespace}{
    \renewcommand{\baselinestretch}{1.6}\large\normalsize}

\newcommand{\bce}{\begin{center}}
\newcommand{\ece}{\end{center}}
\newcommand{\be}{\begin{equation}}
\newcommand{\ee}{\end{equation}}
\newcommand{\bea}{\vspace{0.25cm}\begin{eqnarray}}
\newcommand{\eea}{\end{eqnarray}}

\def\PLA{{Phys. Lett.}  A }
\def\PLB{{Phys. Lett.}  B }
\def\PRL{{Phys. Rev. Lett.} }
\def\PRA{{Phys. Rev.} A }

\def\PRD{{Phys. Rev.} D }

\doublespace

\begin{document}

\title{{\LARGE {\bf On tests of local realism by CP-violation parameters of $K^0$ mesons.}}}

\doublespace

\author{M.Genovese \footnote{ genovese@ien.it. Tel. 39 011 3919253, fax 39 011 3919259}}
\address{Istituto Elettrotecnico Nazionale Galileo Ferraris, Str. delle Cacce
91,\\I-10135 Torino. } \maketitle

\vskip 1cm {\bf Abstract} \vskip 0.5cm

Recently various papers have proposed to test local realism (LR)
by considering electroweak CP-violation parameters values in
neutral pseudoscalar meson systems. Considering the large interest
for a conclusive test of LR and the experimental accessibility to
these tests, in this paper we critically consider these results
showing how they, albeit very interesting, require anyway
additional assumptions and therefore cannot be considered
conclusive tests of LR.
 \vskip 2cm PACS:
13.20.Eb, 03.65.Bz

Keywords: neutral kaons,  Bell inequalities, Pseudoscalar mixing,
non-locality, hidden variable theories

\vskip 0.5cm

{\bf I - Introduction} \vskip 0.5cm
 The problem if a local
realistic theory can reproduce Standard Quantum Mechanics (SQM)
results dates to the early days of this theory. Even if now SQM is
confirmed by a huge amount of data and represents one of the
pillars of modern physics, the fundamental quest about a possible
realistic theory reproducing its results remains open (for a very
recent review see \cite{rev}).

 The earliest discussion about this problem dates to 1935, when Einstein-Podolsky-Rosen
\cite{EPR} suggested that Quantum Mechanics could be an incomplete
theory, representing a statistical approximation of a complete
deterministic theory in which observable values are fixed by some
hidden variable.

In 1964 Bell \cite{Bell} discovered that any realistic Local
Hidden Variable theory (LHVT) must satisfy certain inequalities
which can be violated in SQM leading to the possibility of an
experimental test of the validity of SQM as compared to LHVT.

We would like to emphasize that the great beauty of this theorem
resides in demonstrating in complete generality that every local
realistic theory cannot reproduce all the results of SQM and
therefore one can exclude a whole class of theories with a single
experiment: no other hypothesis beyond locality (namely the
request of no superluminal "connection" between subsystems) is
introduced \footnote{On the other hand, Bell theorem does not
concern non-local hidden variables theories. For a recent progress
toward an experimental verification of one of them see \cite{dbb}
and Ref.s therein.}. Considering the extreme generality of this
theoretical result, one must therefore be careful not to introduce
additional hypotheses when implementing an experiment for testing
it.

Since then, many experiments (mainly  based on entangled photon
pairs) have been addressed to test Bell inequalities
\CITE{asp,pdc,pdc2}, leading to a substantial agreement with
standard quantum mechanics  and strongly disfavouring LHVT. The
request of having well space-like separated measurements
(necessary for excluding every direct influence of measurements
results \footnote{In discussing a completely general {\bf Local}
Hidden Variable Theory one envisages whatever influence between
subsystems with the only condition that it is not superluminal
\cite{Bell,rev}. When discussing a completely general {\bf
Non-Local} Hidden Variable Theory this condition is further
relaxed to the request that eventual superluminal influence among
subsystems does not allow faster than light information
transmission \cite{rev}.}, the so called {\it locality loophole})
has been well verified \cite{asp,pdc2}. However, so far, no
experiment has yet been able to exclude definitively such
theories, since one has always been forced to introduce a further
additional hypothesis \CITE{santos,garuccio}, due to the low total
detection efficiency, stating that the observed sample of particle
pairs is a faithful subsample of the whole ({\it detection or
efficiency loophole}) \footnote{A recent experiment \cite{Win}
based on the use of Be ions has reached very high efficiencies
(around 98 \%), but in this case the two subsystems (the two ions)
are not really separated systems during the measurement and the
test cannot be considered a real implementation of a detection
loophole free test of Bell inequalities, even if it could
represent a progress in this sense.}.  It has been shown that this
loophole could be eliminated only by reaching a detection
efficiency of $0.8284$ when using maximally entangled states or
$0.67$ for non-maximally entangled ones \cite{eb}. Thus, the quest
for new experimental configurations able to overcome the detection
loophole is of
 of the utmost interest.

Recently, many papers \cite{BK} have been devoted to study the
possibility of realizing a conclusive test by the use of
pseudoscalar meson pairs as $K^0 \bar{K^0}$ or $B^0 \bar{B^0}$. If
the pair is produced by the decay of a particle at rest in the
laboratory frame (as the $\phi$ at $Da \phi ne$), the two
particles can be easily separated to a relatively large distance
allowing a space-like separation of the two subsystems and
permitting an easy elimination of the locality loophole, i.e.
realizing two completely space-like separated measurements on the
two subsystems (where the space-like separation must include the
setting of the experimental apparatuses too). A very low noise is
expected as well.

The idea is to use entangled states (i.e. non factorizable in
single particle states) of the form: \bea |\Psi \rangle = { | K^0
\rangle | \bar K^0 \rangle  - | \bar K^0 \rangle | K^0 \rangle
\over \sqrt{2} } = & \cr = { | K_L \rangle | K_S \rangle  - |  K_S
\rangle | K_L \rangle \over \sqrt{2} } & \cr \label{psik} \eea

where $| K^0 \rangle$ and $| \bar K^0 \rangle$ are the particle
and antiparticle related by charge conjugation and composed by a
quark $d$ with an anti-strange $\bar s$ and a $\bar d $ with a $s$
respectively. Whilst mass eigenstates are \be | K_L \rangle = {p |
K^0 \rangle  + q | \bar K^0 \rangle \over \sqrt{|p|^2 + |q|^2}}
\ee and \be | K_S \rangle = { p | K^0 \rangle - q |  \bar K^0
\rangle \over \sqrt{|p|^2 + |q|^2 } } \ee where $p = 1+
\varepsilon$ and $p = 1 - \varepsilon$ in terms of the (small)
 electroweek CP-violation parameter $\varepsilon$ ($|\varepsilon|= (2.26 \pm 0.02) 10^{-3}$).
The $K_L$ is the long living state, corresponding for
$\varepsilon=0$ to $CP=-e^{i \alpha}$ eigenstate ($|K^0_-
\rangle$) for which 2 pions decay is forbidden, and the $K_S$ is
the short living state, corresponding for $\varepsilon=0$ to
$CP=+e^{i \alpha}$ eigenstate ($|K^0_+ \rangle$), for which 2
pions decay is allowed.

Claims that these experimental set-ups \cite{BK} could allow the
elimination of the detection loophole in view of the high
efficiency of particles detectors, have also been made. However,
we have shown \cite{nosK} that due to the necessity of identifying
the state through some of its decays (or interaction) and since
the decay channel (and interaction) can depend on the value of the
hidden variables, the detection loophole appears in this case as
well.

The very recent experimental results of Ref. \cite{belle}
(following some previous ones concerning kaons of Ref.
\cite{CLEO}), with entangled pairs of $B^0 \bar{B}^0$ mesons from
$\Upsilon (4S)$ decay, giving a violation $S = 2.725 \pm
0.167_{stat} \pm 0.092_{syst}$ of CHSH inequality $S < 0$,  are
therefore very interesting representing a test of local realism
for a new kind of particles, but could not lead to an ultimate
test.

In some other recent papers \cite{uch,BGH,BF1,BF2} it has been
studied the connection between the value of CP violating
parameters $\varepsilon$ and $\varepsilon '$ to some  Bell
inequality with the purpose to show that the simple observation of
some specific values of these parameters represent by itself a
test of local realism.

Considering the large relevance of these results,   in this paper
we will carefully analyze the explicit and implicit hypotheses on
which these inequalities are based showing that they do not
represent a clear test of local realism, since other additional
assumptions are needed. The results of Ref. \cite{uch,BGH,BF1,BF2}
are therefore very interesting for having pointed out a connection
between CP-violating parameters and local realism, but do not
allow a conclusive test of this last. \vskip 0.5cm

{\bf II - Discussion of the connection between CP- violation
parameters and local realism.} \vskip 0.5cm

Let us begin by considering the proposal of Ref.s  \cite{BF1,BF2}.

The main idea of these papers is considering a Clauser-Horne like
inequality (one of the many different forms of Bell inequalities
\cite{rev}) on the joint probabilities of observing at a certain
time $t_1$ the first K in a state $f_1$ ($-$ means no selection)
and the second at $t_2$ in a state $f_2$ \bea P(f_1,t_1;f_2,t_2)-
P(f_1,t_1;f_4,t_2)+ P(f_3,t_1;f_2,t_2)+ P(f_3,t_1;f_4,t_2) \nonumber \\
\leq P(f_3,t_1; -,t_2) + P(-,t_1;f_2,t_2) \label{CHK}\eea
afterwards the relation \ref{CHK} is transformed in an inequality
on the CP-violation parameter $\varepsilon '$ (see for example
\cite{PrLe} for its definition), \be |Re\{\varepsilon ' \}| \leq 3
|\varepsilon '|^2 \label{eps'}\ee

In Ref. \cite{BF2} is then shown how the data of KTev \cite{KTev}
and NA48 \cite{NA48} collaborations (obtained with uncorrelated
kaons) violate this inequality of some standard deviations.

However, a first caveat, already indicated in the paper, is that
the demonstration is obtained with the assumption that the decay
of one kaon is stochastically independent from the decay of the
entangled one. In our opinion, this represents by itself a strong
reduction of the generality of LHVT tested by this scheme, since
obviously in a deterministic theory also the decay channel can be
fixed a priori by hidden variables.

However, we would like to emphasize that even larger loopholes
remain in this kind of test of local realism.

First of all, it must be noticed that in a general LHVT framework
the values extracted for $\varepsilon '$ in uncorrelated kaons
decay do not necessarily correspond to what one would obtain with
correlated ones (this of course could eventually be tested in the
future). Furthermore, by using the uncorrelated kaons results of
Ref.s \cite{KTev,NA48} one does not cope at all the problem of
having space-like separated measurements in order to avoid a
possible influence of results.

None the less, even more deep conceptual problems remain unsolved.
In particular, we wish to point out that a further strong
assumption appears, since, in transforming inequality \ref{CHK} in
\ref{eps'}, explicit values of SQM amplitudes are used (including
also amplitudes for the unphysical, being CP-symmetry broken,
$K^0_{\pm}$ states): these relations in principle could not be
valid in a general LHVT. When building a suitable LHV theory one
must satisfy the condition of reproducing SQM results in the sense
of reproducing the good agreement of SQM predictions with {\bf
observed physical quantities} (within nowadays uncertainties), but
no condition is posed for unobserved processes.

In detail, in Ref.s \cite{BF1,BF2} the inequality \ref{CHK} is
rewritten for the probabilities $P(\pi^- l^+ \nu,t; 2 \pi^0,t)$,
$P(\pi^- l^+ \nu,t; 2 \pi^+ \pi^-,t)$ and $P(2 \pi^0 ,t; \pi^+
\pi^- ,t)$ that are
 expressed by using quantum mechanics \cite{3,5} in terms of
the amplitudes $r_{00} $ (the ratio of the decay amplitudes of
$K_+$ and $K_-$ into $\pi^0 \pi^0$), $r_{+-} $ (the ratio of the
decay amplitudes of $K_+$ and $K_-$ into $\pi^+ \pi^-$), x
(describing the violation of $\Delta S = \Delta Q$ rule) and the
parameters $\gamma_S,\gamma_L,\lambda_L$ characterizing the
eigenvalues of effective Hamiltonian decribing $K^0$ time
evolution (see Ref. \cite{BF1,BF2} for definitions): \bea P(\pi^-
l^+ \nu,t; 2 \pi^0,t) = {1 \over 4} e^{-(\gamma_L + \gamma_S) t}
[1
- 2 Re(r_{00})- 2 Re(x)] \nonumber \\
P(\pi^- l^+ \nu,t; 2 \pi^+ \pi^-,t) = {1 \over 4} e^{-(\lambda_L +
\gamma_S)t} [1 - 2 Re(r_{+-})- 2 Re(x)]\\
P(2 \pi^0 ,t; \pi^+ \pi^- ,t) = {1 \over 2} e^{-(\gamma_L +
\gamma_S)t} |r_{+-}-r_{00}|^2 \nonumber \eea

and then in terms of $\varepsilon '$ by using $r_{+-}= \varepsilon
-\epsilon_L + \varepsilon'$ and $r_{00}= \varepsilon -\epsilon_L -
2 \varepsilon'$, where $\varepsilon$ and $\varepsilon '$ signal CP
and CPT-violating effects and can be related to the ratios of
decay amplitudes of $K_{L,S}$ in 2 pions through $\varepsilon +
\varepsilon ' ={A(K_L \rightarrow \pi^+ \pi^-) \over A(K_S
\rightarrow \pi^+ \pi^-) }$ and $\varepsilon - 2 \varepsilon '
={A(K_L \rightarrow \pi^0 \pi^0) \over A(K_S \rightarrow \pi^0
\pi^0) }$ , (see again \cite{BF1,BF2} for a definition of
$\epsilon_L$).

All these relations are calculated in SQM, but in a general LHVT
they do not need to have the same form. When one wants to discuss
in complete generality, like Bell inequalities,  the possible
existence of a local realistic theory, one  must reproduce, within
available uncertainties, the observed quantities as branching
ratios, but all the other amplitudes or parameters and relations
among them, specific of quantum mechanics, cannot be assumed.
Thus, the results of Ref.s \cite{BF1,BF2}, albeit representing an
interesting connection between local realism and specific
properties (CP violation) of electroweak lagrangian, do not lead
to a resolutive test of local realism.

In this sense also the argumentation proposed in Ref. \cite{BF2}
as an indication about the difficulty of building a LHVT where the
two entangled kaons decays are not stochastically independent is
based on the use of quantum evolution and relations among quantum
amplitudes and therefore, for the previous argument, is rather
weak.

For the sake of completeness, let us also notice that a direct use
of inequality \ref{CHK} for joint probabilities measured with an
entangled state \ref{psik} would unavoidably lead to the problem
of a finite efficiency in detecting a certain state. Exactly as
for the other Bell inequalities tests (with mesons \cite{nosK},
photons \cite{santos,garuccio}, etc.)  this requires an additional
hypothesis stating that the selected sample is a faithful
representative of the whole if the total detection efficiency does
not exceed the $82.84 \%$, a condition difficult, if not
impossible, to be realized experimentally.

Then, we discuss  the considerations expressed in Ref. \cite{BGH}.
This paper reconsiders the  original proposal of \cite{uch} to
build a Bell inequality based on joint measurement probabilities
of a $K_S^0$, $\bar K^0$ and the (unphysical) $K^0_+$  with the
state \ref{psik}
 \be P(K_S^0,\bar K^0) \leq P(K_S^0,\bar K^0_+) +
P(K_+^0,\bar K^0) \label{bell}\ee
 and, with
some additional hypothesis on phases, to transform it in an
inequality on the parameter $\varepsilon$, \be Re \{\varepsilon \}
\leq |\varepsilon|^2 \label{eps} \ee

Inequality violated by present data on $\varepsilon$ \cite{PDB},
$\varepsilon = (2.284 \pm 0.014) 10^{-3} e^{i( 43^o.52 \pm 0.06)}
$.

The main idea of Ref. \cite{BGH} is to obtain an inequality
independent by any phase convention. In order to derive this
result Eq. \ref{bell} is rewritten by using the SQM probabilities:
\bea P(K_S^0,\bar K^0) =
{|p|^2 \over 2 \sqrt{|p|^2 + |q|^2 }} \nonumber \\
P(K_S^0,\bar K^0_+)={|p e^{i \alpha}-q|^2 \over  4 \sqrt{|p|^2 +
|q|^2 } } \label{kqp} \\
P(K_+^0,\bar K^0) = 1/4 \nonumber  \eea in the form \be |p| \leq
|q| \label{lpq}\ee

The values of $|p| $ and $|q|$ extracted from semileptonic decays
\cite{PDB} violate this inequality.

Furthermore, by replacing $\bar K^0$ with $K^0$ in the Bell
inequality \ref{bell} one arrives to the inequlity \be |p| \leq
|q| \label{lpq2}\ee that together with \ref{lpq} implies $|p| =
|q|$, in contradiction with experiment \cite{PDB}.

 However, the result \ref{lpq} is not free from
additional assumptions beyond local realism.

Again the relations \ref{kqp} between probabilities and parameters
$p,q$ are based on quantum mechanics (and two of them concern the
unphysical state $K^0_+$). The same considerations about the
criticality of this point for a LHVT discussed for the previous
case remain valid.

Furthermore, the values of $|p|$ and $|q|$ are extracted by
analyzing specific decays of $K^0_{L,S}$. Again, if these values
are not obtained by using space-like separated measurements on
entangled kaons, the locality loophole is not coped. Also, if they
are extracted by specific decay channels, one can conceive LHVT
where the hidden variables determines also the kind of decay and
therefore the values of $|p|$ and $|q|$ determined in some
specific decay are not general parameters pertaining an unbiased
hidden variable sample, which should be used for inequality
\ref{lpq}.

Thus  from all these considerations follows that also the proposal
of Ref. \cite{BGH} is not free from loopholes.

\vskip 0.5cm

{\bf III - Conclusions} \vskip 0.5cm

In summary the results of Ref. \cite{uch,BGH,BF1,BF2} represent an
interesting connection between local realism and electroweak
CP-violation, but are not model independent. Further assumptions
are needed for obtaining these results beyond local realism and
thus they cannot absolutely represent a conclusive test of this
hypothesis, albeit giving interesting indications disfavouring
LHVT. Furthermore, it must be noticed that, in this case, these
additional assumptions are contained in the derivation of the
relations \ref{eps'}, \ref{eps},\ref{lpq}. Therefore, at variance
with the case of detection loophole for photon experiments
\cite{pdc} or locality loophole for ions \cite{Win} that depend on
technological limitations, it will not possible to overcome this
problem.

\vskip 0.5cm

{\bf Acknowledgments}

We acknowledge support  of MIUR (FIRB RBAU01L5AZ-0021) and Regione
Piemonte.

\end{document}